\UseRawInputEncoding
\documentclass[12pt]{article}

\usepackage{amssymb,amsmath,amsthm,amscd,mathrsfs,amsbsy,amsfonts}
\usepackage{pstricks,pst-node}
\usepackage{amsbsy,young,colortbl,cite}

\usepackage{graphicx}

\usepackage{float}

\usepackage{setspace}

\usepackage{ragged2e}

\usepackage{caption}

\usepackage{amsmath}

\usepackage{subfigure}

\usepackage{empheq}

\usepackage{amssymb,amsthm}
\usepackage[english]{babel}
\newtheorem{proposition}{Proposition}[section]

\newtheorem{theorem}[proposition]{Theorem}

\def\({\left(}
\def\){\right)}
\def\[{\begin{eqnarray}}
\def\]{\end{eqnarray}}

\usepackage{amsmath}
\numberwithin{equation}{section}

\begin{document}
\captionsetup[figure]{name={Fig.}}
\title{Darboux transformations and solutions of nonlocal Hirota and Maxwell-Bloch equations
}

\author{
\  \  Ling An\dag, Chuanzhong Li\dag\ddag\footnote{Corresponding author:lichuanzhong@nbu.edu.cn}, Lixiang Zhang\dag\\[4pt]
\small \dag School of Mathematics and Statistics,  Ningbo University, Ningbo, 315211, China\\
\small \ddag College of Mathematics and Systems Science, Shandong University of Science and Technology,\\
\small
 Qingdao, 266590,  China\\}

\date{}

\maketitle

\abstract{In this paper, based on the Hirota and Maxwell-Bloch (H-MB) system and its application in the theory of the femtosecond pulse propagation through an erbium doped fiber, we define two kinds of nonlocal Hirota and Maxwell-Bloch (NH-MB) systems, namely, $PT$-symmetric NH-MB system and reverse space-time NH-MB system. Then we construct the Darboux transformations of these NH-MB systems. Meanwhile, we derive the explicit solutions by the Darboux transformations.}\\

\allowdisplaybreaks
\setlength{\baselineskip}{12pt}
\tableofcontents

\section{Introduction}
\ \ \ \ For a long time, it is difficult and tedious to find the interaction solutions of the nonlinear systems, but it is also meaningful and important. In order to solve the problems of the nonlinear systems, we introduce the symmetry theory\cite{1}, which plays an important role in both integrable and non-integrable systems. The theory of symmetries involves a wide range of contents. Recently, many scholars pay more attention to the nonlocal symmetry, which is proposed by Vinogradov and Krasil¡¯shchik in $1980$\cite{2}. And some studies have shown that the nonlocal symmetry method\cite{3,4,5} is one of the best tools for solving the nonlinear systems. The nonlocal symmetry method is to establish a relationship between the local equations and their corresponding nonlocal equations by selecting appropriate symmetry, so as to study their properties and solutions. Due to the forms of the symmetries are different, there are great differences in the coupling of time and space between these local and nonlocal equations. Therefore new physical phenomena may appear and new physical applications may be produced.

Here, we use AKNS system as an example to explain the origin of nonlocal equations in detail. We all know the soliton equations have many peculiar properties\cite{6}, the most fundamental property of them is that they all can be represented by the integrability conditions of a pair of linear problems, as shown below
\begin{equation}\label{1.1}
\varphi_{t}=M\varphi,
\end{equation}
\begin{equation}\label{1.2}
\varphi_{x}=N\varphi,
\end{equation}
where $\varphi=(\varphi_{1}, \varphi_{2}, \cdots \varphi_{n})^{\mathrm{T}}$ is a $n$-dimensional column vector, $M$ and $N$ are $n$-order matrices, whose elements depend on the spectral parameters $\lambda$ and the potential $u$.

In order to make the Eqs.~\eqref{1.1} and \eqref{1.2} hold at the same time, $\varphi$ must satisfy the compatibility condition $\varphi_{xt}=\varphi_{tx}$, thus $M$ and $N$ must satisfy
\begin{equation}\label{1.3}
M_{x}-N_{t}+[M,N]=0,
\end{equation}
then Eq.~\eqref{1.3} is called Lie group structural equation or zero curvature equation.

If we choose different $M$ and $N$, we can derive different soliton equations. For AKNS system, we usually take
\begin{equation}\label{1.4}
\tilde{M}=\left(
\begin{array}{cc}
-i\lambda & q(x,t)\\[6pt]
r(x,t) & i\lambda\\
\end{array}
\right),
\ \ \ \ \ \ \tilde{N}=\left(
\begin{array}{cc}
A(x,t)&B(x,t)\\[6pt]
C(x,t)&-A(x,t)\\
\end{array}
\right),
\end{equation}
where $A$, $B$ and $C$ are functions containing spectral parameters $\lambda$, functions $q(x,t),$ $r(x,t)$ and their derivatives.

Recently, a series of new nonlocal equations have been reported and studied. The most distinctive kind of nonlocal equations are the equations with $PT$-symmetry\cite{7,8}. An equation with $PT$-symmetry means that the equation is invariant under the joint action of $PT$-operator\cite{3}. In 2013, Ablowitz and Musslimani proposed the first $PT$-symmetric equation. They chose the relation $r(x,t)=q^{*}(-x,t)$, then they got the $PT$-symmetric nonlinear Schr$\ddot{o}$dinger (NLS) equation\cite{4}
\begin{equation}\label{1.5}
iq_{t}(x,t)=q_{xx}(x,t)-2\sigma q^{2}(x,t)q^{*}(-x,t),\ \ \sigma=\mp1.
\end{equation}
Let's rewrite the above equation appropriately
\begin{equation}\label{1.6}
iq_{t}(x,t)=q_{xx}(x,t)+V[q,x,t]q(x,t),
\end{equation}
where $V[q,x,t]=-2\sigma q(x,t)q^{*}(-x,t)$ is called a self-induced potential of this equation, and the self-induced potential satisfies the condition of $PT$-symmetry
\begin{equation*}
V[q,x,t]=V^{*}[q,-x,t].
\end{equation*}
So we call Eq.~\eqref{1.5} the $PT$-symmetric NLS equation.

In $2016$\cite{9}, Ablowitz and Musslimani introduced the integrable nonlocal nonlinear equations systematically, and summarized some nonlocal symmetric forms of AKNS scattering problem, which can be divided into the types
\begin{equation}\label{1.7}
r(x,t)=\sigma q^{*}(\varsigma_{1}x,\varsigma_{2}t),\ \ r(x,t)=\sigma q(\varsigma_{1}x,\varsigma_{2}t),\ \ \ \sigma,\ \varsigma_{1},\ \varsigma_{2}=\pm 1.
\end{equation}
According to these symmetries, we can derive different nonlocal equations.

In this paper, we will focus on the nonlocal form of the Hirota and the Maxwell-Bloch system. The nonlinear Schr$\ddot{o}$dinger equation is a very important equation, which has many physical applications. In $1985$, Kodama found that the higher-order NLS equation could be reduced to the Hirota equation via an appropriate transformation. The Maxwell equation is the core of the electromagnetic theory. It was proposed in the theory of electromagnetic field dynamics, which was published by British physicist James Clerk Maxwell in $1865$\cite{10}. The Bloch equation is one of the important theoretical foundations of nuclear magnetic resonance, and it is also the theoretical basis for the study of transient phenomena of coherent light. And it was proposed by Bloch, Hansen and Packard in $1946$\cite{11}. With the continuous study of the soliton equations, many scholars begin to focus on the coupling of the equations. For the above equations, if the Maxwell equation and the Bloch equation coupling, then we can obtain the Maxwell-Bloch (MB) system\cite{12}. By combining the NLS equation and the MB system together, then we will get the nonlinear Schr$\ddot{o}$dinger and the Maxwell-Bloch (NLS-MB) system\cite{13,14}. Accordingly, on the basis of the NLS equation, we can choose appropriate self steepening and self frequency effects, then the NLS-MB system can be reduced to a coupling system of the Hirota and the Maxwell-Bloch system\cite{15,16,17}. The forms of the H-MB system are given below
\begin{equation}\label{1.8}
\begin{cases}
E_{x}=\beta(E_{ttt}+6|E|^{2}E_{t})+\frac{i}{2}\alpha(E_{tt}+2|E|^{2}E)+2p,\\[4pt]
p_{t}=2E\eta+2i\omega p,\\[4pt]
\eta_{t}=-(pE^{*}+p^{*}E),
\end{cases}
\end{equation}
among them, $x$ and $t$ represent space and time variables respectively, $E(x,t)$ and $p(x,t)$ are complex variables, $\eta(x,t)$ is a real variable corresponding to the extent of the population inversion, $\alpha,\ \beta$ and $\omega$ are real constants and $\omega$ represents the frequency.

In addition, the methods of solving the integrable equations have been developed in recent years, such as Darboux transformation\cite{18,19,20}, Hirota's direct method\cite{21,22}, inverse scattering method\cite{23,24} and so on. The Darboux transformation is one of the most effective methods to solve such equations. In\cite{15}, the authors found many kinds of solutions of H-MB system by using the Darboux transformation method. Therefore, this paper will also use the Darboux transformation to solve the NH-MB system.

The paper is organized as follows. In section $2$, we obtain standard H-MB system and two kinds of NH-MB systems: $PT$-symmetric NH-MB system and reverse space-time NH-MB system. In section $3$, we discuss the Darboux transformation of $PT$-symmetric NH-MB system, give the concrete forms of the new solutions $E^{'},\ p^{'},\ \eta^{'}$ under the one-fold Darboux transformation, and give the determinant form of the $n$-fold Darboux matrix. In section $4$, the solutions of $PT$-symmetric NH-MB system are discussed. Because of the particularity of $\eta$, i.e. $\eta(x,t)=-\eta(-x,t)$, the seed solutions satisfying the system are obviously less than those of standard H-MB system, so we only take two typical seed solutions. In section $5$, we also discuss the Darboux transformation of the reverse space-time NH-MB system and analyze the solutions of this system. Our conclusions are stated in section $6$.

\section{Zero curvature equation}
\ \ \ \ For the H-MB system which belongs to AKNS system discussed in this paper, similarly, we can take
\begin{equation}\label{2.1}
M=\left(
\begin{array}{cc}
-i\lambda & q(x,t)\\[6pt]
r(x,t) & i\lambda\\
\end{array}
\right),
\ \ \ \ \ \ N=\left(
\begin{array}{cc}
N_{11}(x,t)&N_{12}(x,t)\\[6pt]
N_{21}(x,t)&-N_{11}(x,t)\\
\end{array}
\right).
\end{equation}
Substituting \eqref{2.1} into the compatibility condition \eqref{1.3}, we can obtain
\begin{equation}\label{2.2}
\begin{cases}
N_{11t}(x,t)=q(x,t)N_{21}(x,t)-r(x,t)N_{12}(x,t),\\[3pt]
q_{x}(x,t)=N_{12t}(x,t)+2i\lambda N_{12}(x,t)+2q(x,t)N_{11}(x,t),\\[3pt]
r_{x}(x,t)=N_{21t}(x,t)-2i\lambda N_{21}(x,t)-2r(x,t)N_{11}(x,t).
\end{cases}
\end{equation}
Next, we make
\begin{align*}
N_{11}(x,t)&=4i\beta \lambda^{3}-i\alpha \lambda^{2}+2i\beta qr\lambda +\Big[\beta(qr_{t}-rq_{t})-\frac{i}{2}\alpha qr\Big]+i\eta (\lambda+\omega)^{-1},\\
N_{12}(x,t)&=-4\beta q\lambda^{2}+(-2i\beta q_{t}+\alpha q)\lambda +\Big[\beta(q_{tt}-2q^{2}r)+\frac{i}{2}\alpha q_{t}\Big]-i\delta p(\lambda+\omega)^{-1},\\
N_{21}(x,t)&=-4\beta r\lambda^{2}+(2i\beta r_{t}+\alpha r)\lambda +\Big[\beta(r_{tt}-2qr^{2})-\frac{i}{2}\alpha r_{t}\Big]-i\delta m(\lambda+\omega)^{-1},
\end{align*}
that is to say
\begin{equation}\label{2.3}
\begin{split}
N=&
\left(
\begin{array}{cc}
4i\beta&0\\[6pt]
0&-4i\beta\\
\end{array}
\right)\lambda^{3}+
\left(
\begin{array}{cc}
-i\alpha&-4\beta q\\[6pt]
-4\beta r&i\alpha\\
\end{array}
\right)\lambda^{2}+
\left(
\begin{array}{cc}
2i\beta q r&-2i\beta q_{t}+\alpha q\\[6pt]
2i\beta r_{t}+\alpha r&-2i\beta qr\\
\end{array}
\right)\lambda\\[5pt]
&+
\left(
\begin{array}{cc}
\beta(qr_{t}-rq_{t})-\frac{i}{2}\alpha qr&\beta(q_{tt}-2q^{2}r)+\frac{i}{2}\alpha q_{t}\\[6pt]
\beta(r_{tt}-2qr^{2})-\frac{i}{2}\alpha r_{t}&-\beta(qr_{t}-rq_{t})+\frac{i}{2}\alpha qr\\
\end{array}
\right)+\frac{i}{\lambda+\omega}
\left(
\begin{array}{cc}
\eta&-\delta p\\[6pt]
-\delta m&-\eta\\
\end{array}
\right)\\
:=&N_{3}\lambda^{3}+N_{2}\lambda^{2}+N_{1}\lambda+N_{0}+\frac{i}{\lambda+\omega}N_{-1}.\\[-20pt]
\end{split}
\end{equation}
Plugging the concrete forms of $N_{11}(x,t),$ $N_{12}(x,t)$ and $N_{21}(x,t)$ into \eqref{2.2}, five equalities about functions $q(x,t),\ r(x,t),\ p(x,t),\ m(x,t)\ and\ \eta(x,t)$ can be obtained
\begin{equation}\label{2.4}
\begin{cases}
q_{x}=\beta(q_{ttt}-6qrq_{t})+\frac{i}{2}\alpha(q_{tt}-2q^{2}r)+2\delta p,\\
r_{x}=\beta(r_{ttt}-6qrr_{t})-\frac{i}{2}\alpha(r_{tt}-2qr^{2})-2\delta m,\\
p_{t}=2\delta q\eta+2i\omega p,\\
m_{t}=-2\delta r\eta-2i\omega m,\\
\eta_{t}=\delta(pr-mq).\\
\end{cases}
\end{equation}
If $q(x,t),\ r(x,t),\ p(x,t)\ and\ m(x,t)$ are selected properly, the above equalities can be reduced to three equalities.

\subsection{Standard H-MB system}
\textbf{Standard symmetry: }
\begin{equation}\label{2.5}
r(x,t)=\sigma q^{*}(x,t)=\sigma E^{*}(x,t),\ \ m(x,t)=\delta p^{*}(x,t).
\end{equation}

Taking the above symmetric forms into Eq.~\eqref{2.4}, we can obtain
\begin{equation}\label{2.6}
\begin{cases}
E_{x}=\beta(E_{ttt}-6\sigma EE^{*}E_{t})+\frac{i}{2}\alpha(E_{tt}-2\sigma E^{2}E^{*})+2\delta p,\\
\sigma E^{*}_{x}=\sigma\beta(E^{*}_{ttt}-6\sigma EE^{*}E^{*}_{t})-\frac{i}{2}\alpha\sigma\big(E^{*}_{tt}-2\sigma EE^{*2}\big)-2p^{*},\\
p_{t}=2\delta E\eta+2i\omega p,\\
\delta p^{*}_{t}=-2\sigma\delta E^{*}\eta-2i\delta \omega p^{*},\\
\eta_{t}=\sigma\delta pE^{*}-p^{*}E,
\end{cases}
\end{equation}
when $\sigma=-1,\ \delta=1$ and $\alpha,\ \beta,\ \omega\in\mathbb{R},$ there are
\begin{subequations}\label{2.7}
\begin{empheq}[left=\empheqlbrace]{align}
E_{x}&=\beta(E_{ttt}+6EE^{*}E_{t})+\frac{i}{2}\alpha(E_{tt}+2E^{2}E^{*})+2p,\\
E^{*}_{x}&=\beta(E^{*}_{ttt}+6EE^{*}E^{*}_{t})-\frac{i}{2}\alpha\big(E^{*}_{tt}+2EE^{*2}\big)+2p^{*},\\
p_{t}&=2E\eta+2i\omega p,\\[3pt]
p^{*}_{t}&=2E^{*}\eta-2i\omega p^{*},\\[3pt]
\eta_{t}&=-(pE^{*}+p^{*}E).
\end{empheq}
\end{subequations}
By taking the conjugate on both sides of equality $(2.7b)$, we can get $(2.7a)$. Similarly, the equality $(2.7d)$ can be reduced to $(2.7c),$ then the equalities $(2.7a)$, $(2.7c)$ and $(2.7e)$ constitute the H-MB system \eqref{1.8}.

When $\alpha=1,\ \beta=0$, the H-MB system can be converted to the NLS-MB system
\begin{equation}\label{2.8}
\begin{cases}
E_{x}=\frac{i}{2}\big(E_{tt}+2|E|^{2}E\big)+2p,\\[4pt]
p_{t}=2E\eta+2i\omega p,\\[4pt]
\eta_{t}=-(pE^{*}+p^{*}E).
\end{cases}
\end{equation}

When $\alpha=0,\ \beta=1$, the H-MB system can be converted to the mKdV-MB system
\begin{equation}\label{2.9}
\begin{cases}
E_{x}=E_{ttt}+6|E|^{2}E_{t}+2p,\\[4pt]
p_{t}=2E\eta+2i\omega p,\\[4pt]
\eta_{t}=-(pE^{*}+p^{*}E).
\end{cases}
\end{equation}

\subsection{Nonlocal H-MB system}
\subsubsection*{2.2.1 $PT$-symmetric NH-MB system}
\textbf{$PT$-symmetry: }
\begin{equation}\label{2.10}
r(x,t)=\sigma q^{*}(-x,t)=\sigma E^{*}(-x,t),\ \ m(x,t)=\delta p^{*}(-x,t).
\end{equation}

Substituting the above symmetric forms into Eq.~\eqref{2.4}, we consider the first and second equalities of Eq.~\eqref{2.4}, there are
\begin{equation}\label{2.11}
\begin{split}
E_{x}(x,t)=&\beta(E_{ttt}(x,t)-6\sigma E(x,t)E^{*}(-x,t)E_{t}(x,t))\\
&+\frac{i}{2}\alpha(E_{tt}(x,t)-2\sigma E^{2}(x,t)E^{*}(-x,t))+2\delta p(x,t),
\end{split}
\end{equation}
\begin{equation}\label{2.12}
\begin{split}
\sigma E^{*}_{x}(-x,t)=&\sigma\beta(E^{*}_{ttt}(-x,t)-6\sigma E(x,t)E^{*}(-x,t)E^{*}_{t}(-x,t))\\
&-\frac{i}{2}\alpha\sigma\big(E^{*}_{tt}(-x,t)-2\sigma E(x,t)E^{*2}(-x,t)\big)-2p^{*}(-x,t).
\end{split}
\end{equation}
If we do the variable transformations $x\rightarrow -x$ and $t\rightarrow t$ for Eq.~\eqref{2.12}, and then take the conjugate on both sides of Eq.~\eqref{2.12} at the same time, we get
\begin{equation}\label{2.13}
\begin{split}
E_{x}(x,t)=&-\beta^{*}(E_{ttt}(x,t)-6\sigma E(x,t)E^{*}(-x,t)E_{t}(x,t))\\
&-\frac{i}{2}\alpha^{*}(E_{tt}(x,t)-2\sigma E^{2}(x,t)E^{*}(-x,t))+2\sigma p(x,t).
\end{split}
\end{equation}
To make Eq.~\eqref{2.13} compatible with Eq.~\eqref{2.11}, if and only if $\sigma=\delta,~\beta^{*}=-\beta,~\alpha^{*}=-\alpha,$ that is, $\alpha$ and $\beta$ need to be pure imaginary numbers. Next we consider the third and fourth equalities of Eq.~\eqref{2.4}, which become
\begin{equation}\label{2.14}
p_{t}(x,t)=2\delta E(x,t)\eta(x,t)+2i\omega p(x,t),
\end{equation}
\begin{equation}\label{2.15}
\delta p^{*}_{t}(-x,t)=-2\sigma\delta E^{*}(-x,t)\eta(x,t)-2i\delta \omega p^{*}(-x,t).
\end{equation}
If we do the same transformation for Eq.~\eqref{2.15} as for Eq.~\eqref{2.12}, we get
\begin{equation}\label{2.16}
p_{t}(x,t)=-2\sigma E(x,t)\eta(-x,t)+2i\omega^{*}p(x,t).
\end{equation}
To make Eq.~\eqref{2.16} compatible with Eq.~\eqref{2.14}, if and only if $\omega^{*}=\omega,~\eta(x,t)=-\sigma\delta\eta(-x,t).$ Combining the above discussion, we can get the constraint conditions $\sigma=\delta,~\omega\in\mathbb{R},~\alpha,~\beta,$ are all pure imaginary numbers and $\eta(x,t)=-\eta(-x,t),$ and under these conditions, if we do the same transformation for the fifth equality of Eq.~\eqref{2.4}, we can find that it is compatible with itself. Then we can get the $PT$-symmetric NH-MB system
\begin{equation}\label{2.17}
\begin{cases}
\begin{aligned}
E_{x}(x,t)=&\beta\big(E_{ttt}(x,t)-6\sigma E(x,t)E^{*}(-x,t)E_{t}(x,t)\big)\\
&+\frac{i}{2}\alpha\big(E_{tt}(x,t)-2\sigma E^{2}(x,t)E^{*}(-x,t)\big)+2\sigma p(x,t),
\end{aligned}\\[4pt]
\begin{aligned}
p_{t}(x,t)=&2\sigma E(x,t)\eta(x,t)+2i\omega p(x,t),
\end{aligned}\\[4pt]
\begin{aligned}
\eta_{t}(x,t)=&p(x,t)E^{*}(-x,t)-p^{*}(-x,t)E(x,t),
\end{aligned}
\end{cases}
\end{equation}
where $\alpha=i\varepsilon_{1},\ \beta=i\varepsilon_{2},\ \varepsilon_{1},\ \varepsilon_{2},\ \omega\in\mathbb{R}$, and satisfy the requirement of \begin{equation*}
\eta(x,t)=-\eta(-x,t).
\end{equation*}
Since the $PT$-symmetric NH-MB system \eqref{2.17} comes out of the new symmetry reductions, it is also an infinite dimensional integrable Hamiltonian system. An infinite number of conservation laws of Eq.~\eqref{2.17} can be obtained by expanding the spectral parameter $\lambda$ at the point of infinity. Here, we gave the general formulas of the conserved quantities
\begin{equation}\label{2.18}
\begin{cases}
\begin{aligned}
I_{1}=\int_{-\infty}^{+\infty}\Big[&2\sigma\eta(x,t)-3i\sigma\varepsilon_{2}E^{2}(x,t)E^{*2}(-x,t)+\frac{\varepsilon_{1}}{2}\big(E(x,t)E^{*}_{t}(-x,t)\\
&-E_{t}(x,t)E^{*}(-x,t)\big)+i\varepsilon_{2}\big(E(x,t)E^{*}_{tt}(-x,t)\\
&+E_{tt}(x,t)E^{*}(-x,t)-E_{t}(x,t)E^{*}_{t}(-x,t)\big)\Big]dt,
\end{aligned}
\\[5pt]
\begin{aligned}
I_{n+1}=\int_{-\infty}^{+\infty}\Big[&(-2)^{n+1}(i\omega)^{n}\eta(x,t)+2\sigma p(x,t)\sum\limits_{j=0}^{n-1}\frac{(-2i\omega)^{j}W_{n-j-1}(x,t)}{E(x,t)}+i\varepsilon_{2}W_{n+2}(x,t)\\
&+\big(-i\varepsilon_{2}\frac{E_{t}(x,t)}{E(x,t)}+\frac{\varepsilon_{1}}{2}\big)W_{n+1}(x,t)+\Big(i\varepsilon_{2}\big(\frac{E_{tt}(x,t)}{E(x,t)}-2\sigma E(x,t)E^{*}(-x,t)\big)\\
&-\frac{\varepsilon_{1}E_{t}(x,t)}{2E(x,t)}\Big) W_{n}(x,t)\Big]dt,\ \ \ n\geq1,
\end{aligned}
\end{cases}
\end{equation}
where
\begin{equation*}
\begin{split}
W_{0}(x,t)&=-\sigma E(x,t)E^{*}(-x,t),\ \ \ \ W_{1}(x,t)=-\sigma E(x,t)E^{*}_{t}(-x,t),\\
W_{n}(x,t)&=E(x,t)(\frac{W_{n-1}(x,t)}{E(x,t)})_{t}+\sum\limits_{j=0}^{n-2}W_{j}(x,t)W_{n-j-2}(x,t),\ \ \ n\geq2.
\end{split}
\end{equation*}

\subsubsection*{2.2.2 Reverse space-time NH-MB system}
\textbf{Reverse space-time symmetry: }
\begin{equation}\label{2.19}
r(x,t)=\sigma q(-x,-t)=\sigma E(-x,-t),\ \ m(x,t)=\delta p(-x,-t)=\sigma p(-x,-t).
\end{equation}

Taking the above symmetric forms into Eq.~\eqref{2.4}, and using a similar approach to that of the $PT$-symmetric NH-MB system, we can get the reverse space-time NH-MB system
\begin{equation}\label{2.20}
\begin{cases}
\begin{aligned}
E_{x}(x,t)=&\beta\big(E_{ttt}(x,t)-6\sigma E(x,t)E(-x,-t)E_{t}(x,t)\big)\\
&+\frac{i}{2}\alpha\big(E_{tt}(x,t)-2\sigma E^{2}(x,t)E(-x,-t)\big)+2\sigma p(x,t),
\end{aligned}\\[4pt]
\begin{aligned}
p_{t}(x,t)=2\sigma E(x,t)\eta(x,t)+2i\omega p(x,t),
\end{aligned}\\[4pt]
\begin{aligned}
\eta_{t}(x,t)=p(x,t)E(-x,-t)-p(-x,-t)E(x,t),
\end{aligned}
\end{cases}
\end{equation}
where $\alpha,\ \beta\in\mathbb{C},\ \omega\in\mathbb{R}$, and satisfying the requirement of $\eta(x,t)=\eta(-x,-t)$.
The conserved quantities associated with Eq.~\eqref{2.20} are
\begin{equation}\label{2.21}
\begin{cases}
\begin{aligned}
I_{1}=\int_{-\infty}^{+\infty}&\Big[2\sigma\eta(x,t)-3\sigma\beta E^{2}(x,t)E^{2}(-x,-t)\\
&-\frac{i\alpha}{2}\big(E(x,t)E_{t}(-x,-t)-E_{t}(x,t)E(-x,-t)\big)\\
&+\beta\big(E(x,t)E_{tt}(-x,-t)+E_{tt}(x,t)E(-x,-t)-E_{t}(x,t)E_{t}(-x,-t)\big)\Big]dt,
\end{aligned}
\\[6pt]
\begin{aligned}
I_{n+1}=\int_{-\infty}^{+\infty}&\Big[(-2)^{n+1}(i\omega)^{n}\eta(x,t)+2\sigma p(x,t)\sum\limits_{j=0}^{n-1}\frac{(-2i\omega)^{j}W_{n-j-1}(x,t)}{E(x,t)}\\
&+\beta W_{n+2}(x,t)-\big(\beta\frac{E_{t}(x,t)}{E(x,t)}+\frac{i\alpha}{2}\big)W_{n+1}(x,t)\\
&+\Big(\beta\big(\frac{E_{tt}(x,t)}{E(x,t)}-2\sigma E(x,t)E(-x,-t)\big)+\frac{i\alpha E_{t}(x,t)}{2E(x,t)}\Big) W_{n}(x,t)\Big]dt,\ \ n\geq1,
\end{aligned}
\end{cases}
\end{equation}
where
\begin{equation*}
\begin{split}
W_{0}(x,t)&=-\sigma E(x,t)E(-x,-t),\ \ \ \ W_{1}(x,t)=-\sigma E(x,t)E_{t}(-x,-t),\\
W_{n}(x,t)&=E(x,t)(\frac{W_{n-1}(x,t)}{E(x,t)})_{t}+\sum\limits_{j=0}^{n-2}W_{j}(x,t)W_{n-j-2}(x,t),\ \ \ n\geq2.
\end{split}
\end{equation*}

For the other symmetric forms in \eqref{1.7}, we found that Eq.~\eqref{2.4} can not be successfully coupled through calculation and analysis. Here, we take the symmetric form $r(x,t)=\sigma q^{*}(-x,-t)=\sigma E^{*}(-x,-t),\ \ m(x,t)=\delta p^{*}(-x,-t),$ as an example to illustrate. Substituting the above symmetric forms into Eq.~\eqref{2.4}, there are
\begin{equation}\label{2.22}
\begin{split}
E_{x}(x,t)=&\beta(E_{ttt}(x,t)-6\sigma E(x,t)E^{*}(-x,-t)E_{t}(x,t))\\
&+\frac{i}{2}\alpha(E_{tt}(x,t)-2\sigma E^{2}(x,t)E^{*}(-x,-t))+2\delta p(x,t),
\end{split}
\end{equation}
\begin{equation}\label{2.23}
\begin{split}
\sigma E^{*}_{x}(-x,-t)=&\sigma\beta(E^{*}_{ttt}(-x,-t)-6\sigma E(x,t)E^{*}(-x,-t)E^{*}_{t}(-x,-t))\\
&-\frac{i}{2}\alpha\sigma\big(E^{*}_{tt}(-x,-t)-2\sigma E(x,t)E^{*2}(-x,-t)\big)-2p^{*}(-x,-t).
\end{split}
\end{equation}
\begin{equation}\label{2.24}
p_{t}(x,t)=2\delta E(x,t)\eta(x,t)+2i\omega p(x,t),
\end{equation}
\begin{equation}\label{2.25}
\delta p^{*}_{t}(-x,-t)=-2\sigma\delta E^{*}(-x,-t)\eta(x,t)-2i\delta \omega p^{*}(-x,-t),
\end{equation}
\begin{equation}\label{2.26}
\eta_{t}(x,t)=\sigma\delta p(x,t)E^{*}(-x,-t)-p^{*}(-x,-t)E(x,t).
\end{equation}
If we do the variable transformations $x\rightarrow -x$ and $t\rightarrow -t$ for Eqs.~\eqref{2.23} and \eqref{2.25}, and then take the conjugate on both sides of them, we can obtain
\begin{equation}\label{2.27}
\begin{split}
E_{x}(x,t)=&\beta^{*}(E_{ttt}(x,t)-6\sigma E(x,t)E^{*}(-x,-t)E_{t}(x,t))\\
&-\frac{i}{2}\alpha^{*}(E_{tt}(x,t)-2\sigma E^{2}(x,t)E^{*}(-x,-t))+2\sigma p(x,t).
\end{split}
\end{equation}
\begin{equation}\label{2.28}
p_{t}(x,t)=2\sigma E(x,t)\eta(-x,-t)-2i\omega^{*}p(x,t).
\end{equation}
In order to make Eq.~\eqref{2.22} compatible with Eq.~\eqref{2.27}, it can be found that conditions
\begin{equation*}
\alpha^{*}=-\alpha,\ \ \beta^{*}=\beta,\ \ \sigma\cdot\delta=1,
\end{equation*}
must be satisfied, while conditions
\begin{equation*}
\omega^{*}=-\omega,\ \ \eta(x,t)=\sigma\cdot\delta\eta(-x,-t),
\end{equation*}
must be satisfied when Eq.~\eqref{2.24} compatible with Eq.~\eqref{2.28}. But $\omega$ represents the frequency which must be a real constant. That is, Eq.~\eqref{2.4} cannot be coupled successfully when the symmetric forms $r(x,t)=\sigma q^{*}(-x,-t)=\sigma E^{*}(-x,-t),\ \ m(x,t)=\delta p^{*}(-x,-t)$ are taken. Therefore, there are only two kinds of the NH-MB system.

\section{Darboux transformation of $PT$-symmetric NH-MB system}
\ \ \ \ Based on the Darboux transformation of AKNS system, we will study the Darboux transformation of $PT$-symmetric NH-MB system \eqref{2.17} in this section. First we consider a similar gauge transformation
\begin{equation}\label{3.1}
\varphi^{'}=T\varphi=(\lambda A-S)\varphi,
\end{equation}
among them,
\begin{equation}\label{3.2}
A=\left(
\begin{array}{cc}
a_{11} & a_{12}\\[6pt]
a_{21} & a_{22}\\
\end{array}
\right),
\ \ \ \ S=\left(
\begin{array}{cc}
s_{11} & s_{12}\\[6pt]
s_{21} & s_{22}\\
\end{array}
\right).
\end{equation}

Assume that the new function $\varphi^{'}$ satisfies the equations
\begin{equation}\label{3.3}
\varphi^{'}_{t}=M^{'}\varphi^{'},\ \ \ \ \varphi^{'}_{x}=N^{'}\varphi^{'},
\end{equation}
where $M^{'},\ N^{'}$ depends on $E^{'},\ p^{'},\ \eta^{'},$ and the relationships between $M^{'},\ N^{'}$ and $E^{'},\ p^{'},\ \eta^{'}$ are the same as those between $M,\ N$ and $E,\ p,\ \eta$.

We can also assume that the function $\varphi^{'}$ satisfies the equation
\begin{equation}\label{3.4}
\varphi^{'}_{tx}=\varphi^{'}_{xt}.
\end{equation}
Let's substitute \eqref{3.1} into \eqref{3.3}, combined with \eqref{1.1} and \eqref{1.2}, we can get
\begin{equation}\label{3.5}
T_{t}=M^{'}T-TM,
\end{equation}
\begin{equation}\label{3.6}
T_{x}=N^{'}T-TN,
\end{equation}
next we will discuss the Eqs.~\eqref{3.5} and \eqref{3.6} respectively.

Firstly, we substitute the forms of $A$ and $S$ into \eqref{3.5} to simplify it, combine the symmetric form \eqref{2.10} involved in this system and compare the coefficients of each power of $\lambda$, we can find that when $\sigma=1$, there are
\begin{equation*}
s_{11}(x,t)=\hat{\xi}s_{22}^{*}(-x,t),\ s_{12}(x,t)=\hat{\xi}s_{21}^{*}(-x,t),\ a_{11}(x,t)=\hat{\xi}a_{22}^{*}(-x,t),\ \ \hat{\xi}=\pm 1;
\end{equation*}
when $\sigma=-1$, there are
\begin{equation*}
s_{11}(x,t)=\check{\xi}s_{22}^{*}(-x,t),\ s_{12}(x,t)=-\check{\xi}s_{21}^{*}(-x,t),\ a_{11}(x,t)=\check{\xi}a_{22}^{*}(-x,t),\ \ \check{\xi}=\pm 1.
\end{equation*}
Then we can obtain
\begin{equation}\label{3.7}
a_{22}E^{'}=a_{11}E-2is_{12},
\end{equation}
\begin{equation*}
a_{12}=a_{21}=0,\ \ a_{11t}=a_{22t}=0,
\end{equation*}
\begin{equation*}
S_{t}=M^{'}_{0}S-SM_{0}=[M_{0},S]+i[S,\sigma_{3}]S,\ \ \
\end{equation*}
where
\begin{equation*}
M_{0}=\left(
\begin{array}{cc}
0&E(x,t)\\[6pt]
\sigma E^{*}(-x,t)&0\\
\end{array}
\right),
\ \sigma_{3}=\left(
\begin{array}{cc}
1&0\\[6pt]
0&-1\\
\end{array}
\right),
\ M^{'}_{0}=\left(
\begin{array}{cc}
0&E^{'}(x,t)\\[6pt]
\sigma E^{*'}(-x,t)&0\\
\end{array}
\right).
\end{equation*}

After the same calculations for Eq.~\eqref{3.6}, we find that the conclusions are consistent with the above conclusions, and we get
\begin{equation}\label{3.8}
N^{'}_{-1}=(S+\omega A)N_{-1}(S+\omega A)^{-1},\ \ \ \
\end{equation}
\begin{equation*}
S_{x}=N^{'}_{0}S-SN_{0}-i(N^{'}_{-1}A-AN_{-1}).
\end{equation*}
In order to facilitate the subsequent calculations and analyses, we take $A=I$ here, so
\begin{align*}
\sigma=1:&\  s_{11}(x,t)=s_{22}^{*}(-x,t),\ s_{12}(x,t)=s_{21}^{*}(-x,t);\\
\sigma=-1:&\  s_{11}(x,t)=s_{22}^{*}(-x,t),\ s_{12}(x,t)=-s_{21}^{*}(-x,t).
\end{align*}

\subsection{One-fold Darboux transformation}

\ \ \ \ We assume
\begin{equation}\label{3.9}
S=H\wedge H^{-1},
\end{equation}
where
\begin{equation*}
\wedge=\left(
\begin{array}{cc}
\lambda_{1} & 0\\[6pt]
0 & \lambda_{2}\\
\end{array}
\right),
\end{equation*}
\begin{equation*}
H=
\left(
\begin{array}{cc}
\varphi_{1}(x,t,\lambda_{1})&\varphi_{1}(x,t,\lambda_{2})\\[6pt]
\varphi_{2}(x,t,\lambda_{1})&\varphi_{2}(x,t,\lambda_{2})\\
\end{array}
\right)
:=\left(
\begin{array}{cc}
\varphi_{1,1}(x,t)&\varphi_{1,2}(x,t)\\[6pt]
\varphi_{2,1}(x,t)&\varphi_{2,2}(x,t)\\
\end{array}
\right).
\end{equation*}
According to the symmetric form \eqref{2.10} involved in this system and combing with the relationships between $s_{11},\ s_{12}$ and $s_{21},\ s_{22}$, there are $\lambda_{2}=\lambda^{*}_{1},$ and
\begin{align*}
\sigma=1:&\  \varphi_{1,2}(x,t)=\hat{\zeta}\varphi_{2,1}^{*}(-x,t),\  \varphi_{2,2}(x,t)=\hat{\zeta}\varphi_{1,1}^{*}(-x,t),\ \ \hat{\zeta}=\pm 1;\\
\sigma=-1:&\  \varphi_{1,2}(x,t)=\check{\zeta}\varphi_{2,1}^{*}(-x,t),\  \varphi_{2,2}(x,t)=-\check{\zeta}\varphi_{1,1}^{*}(-x,t),\ \ \check{\zeta}=\pm 1.
\end{align*}
Therefore
\begin{equation}\label{3.10}
S=\frac{1}{\Delta}
\cdot\left(
\begin{array}{cc}
\lambda_{1}\varphi_{1,1}\varphi_{1,1}^{*}(-x,t)-\sigma\lambda_{1}^{*}\varphi_{2,1}\varphi_{2,1}^{*}(-x,t)&-\sigma(\lambda_{1}-\lambda_{1}^{*})\varphi_{1,1}\varphi_{2,1}^{*}(-x,t)\\[6pt]
(\lambda_{1}-\lambda_{1}^{*})\varphi_{1,1}^{*}(-x,t)\varphi_{2,1}&\lambda_{1}^{*}\varphi_{1,1}\varphi_{1,1}^{*}(-x,t)-\sigma\lambda_{1}\varphi_{2,1}\varphi_{2,1}^{*}(-x,t)\\
\end{array}
\right),\\[4pt]
\end{equation}
where $\Delta=\varphi_{1,1}(x,t)\varphi_{1,1}^{*}(-x,t)-\sigma\varphi_{2,1}(x,t)\varphi_{2,1}^{*}(-x,t)$.

Then we substitute \eqref{3.10} into \eqref{3.7} and \eqref{3.8}, the relationships between the new solutions $E^{'},\ p^{'},\ \eta^{'}$ and the old solutions $E,\ p,\ \eta$ of Eq.~\eqref{2.17} can be obtained
\begin{equation}\label{3.11}
E^{'}(x,t)=E(x,t)+\frac{2\sigma i(\lambda_{1}-\lambda^{*}_{1})\varphi_{1,1}(x,t)\varphi_{2,1}^{*}(-x,t)}{\varphi_{1,1}(x,t)\varphi_{1,1}^{*}(-x,t)-\sigma\varphi_{2,1}(x,t)\varphi_{2,1}^{*}(-x,t)},\ \ \ \ \ \ \ \ \ \ \ \ \ \ \ \ \ \ \ \ \ \ \
\end{equation}
\begin{equation}\label{3.12}
\begin{split}
p^{'}(x,t)=&\frac{1}{\Delta^{'}}
\Big[-\sigma p^{*}(-x,t)\cdot\big((\lambda_{1}-\lambda_{1}^{*})\varphi_{1,1}(x,t)\varphi_{2,1}^{*}(-x,t)\big)^{2}\\[4pt]
&+p(x,t)\cdot\big((\omega+\lambda_{1})\varphi_{1,1}(x,t)\varphi_{1,1}^{*}(-x,t)-\sigma(\omega+\lambda_{1}^{*})\varphi_{2,1}(x,t)
\varphi_{2,1}^{*}(-x,t)\big)^{2}\\[4pt]
&-2\eta(x,t)\cdot
(\lambda_{1}-\lambda_{1}^{*})\varphi_{1,1}(x,t)\varphi_{2,1}^{*}(-x,t)\cdot\big((\omega+\lambda_{1})\varphi_{1,1}(x,t)\varphi_{1,1}^{*}(-x,t)\\[4pt]
&-\sigma(\omega+\lambda_{1}^{*})\varphi_{2,1}(x,t)\varphi_{2,1}^{*}(-x,t)\big)\Big],\\[-11pt]
\end{split}
\end{equation}
\begin{equation}\label{3.13}
\begin{split}
\eta^{'}(x,t)=&\frac{\sigma}{\Delta^{'}}
\Big[p^{*}(-x,t)\cdot(\lambda_{1}-\lambda_{1}^{*})\varphi_{1,1}(x,t)\varphi_{2,1}^{*}(-x,t)\\[4pt]
&\cdot\big((\omega+\lambda_{1}^{*})\varphi_{1,1}(x,t)\varphi_{1,1}^{*}(-x,t)-\sigma(\omega+\lambda_{1})\varphi_{2,1}(x,t)
\varphi_{2,1}^{*}(-x,t)\big)\\[4pt]
&+p(x,t)\cdot(\lambda_{1}-\lambda_{1}^{*})\varphi_{1,1}^{*}(-x,t)\varphi_{2,1}(x,t)\\[4pt]
&\cdot\big((\omega+\lambda_{1})\varphi_{1,1}(x,t)\varphi_{1,1}^{*}(-x,t)-\sigma(\omega+\lambda_{1}^{*})\varphi_{2,1}(x,t)
\varphi_{2,1}^{*}(-x,t)\big)\Big]\\[4pt]
&+\eta(x,t)\cdot\big(1-\frac{2}{\Delta^{'}}(\lambda_{1}-\lambda_{1}^{*})^{2}\varphi_{1,1}(x,t)\varphi_{1,1}^{*}(-x,t)\varphi_{2,1}(x,t)\varphi_{2,1}^{*}(-x,t)            \big),\\
\end{split}
\end{equation}
where
\begin{equation*}
\Delta^{'}=(\omega+\lambda_{1})(\omega+\lambda_{1}^{*})\big(\varphi_{11}(x,t)\varphi_{11}^{*}(-x,-t)-\sigma\varphi_{21}(x,t)\varphi_{21}^{*}(-x,-t)\big)^{2}.
\end{equation*}

\subsection{$N$-fold Darboux transformation}

\ \ \ \
We found that the forms of the solutions of $PT$-symmetric NH-MB system \eqref{2.17} obtained by using the one-fold Darboux transformation were already very cumbersome, so we introduced the determinant representations of Darboux transformation to represent the solutions of Eq.~\eqref{2.17}.
\begin{theorem}
For the $n$-fold Darboux transformation of the $PT$-symmetric NH-MB system \eqref{2.17}, its corresponding Darboux matrix form is as follows
\begin{equation}\label{3.14}
\begin{split}
T_{n}(\lambda;\lambda_{1},\lambda_{2},\cdots,\lambda_{2n})&=\lambda^{n}I-S^{[n]}_{n-1}\lambda^{n-1}-\cdots-S^{[n]}_{1}\lambda-S^{[n]}_{0}\\
&=\frac{1}{|(H^{[n]})^{\mathrm{T}}|}
\left(
\begin{array}{cc}
|\mathbb{T}^{[n]}_{11}| & |\mathbb{T}^{[n]}_{12}|\\[5pt]
|\mathbb{T}^{[n]}_{21}| & |\mathbb{T}^{[n]}_{22}|\\
\end{array}
\right):=
\left(
\begin{array}{cc}
(T_{n})_{11} & (T_{n})_{12}\\[5pt]
(T_{n})_{21} & (T_{n})_{22}\\
\end{array}
\right),
\end{split}
\end{equation}
where $\lambda^{*}_{2n-1}=\lambda_{2n}$,
\begin{align*}
\mathbb{T}^{[n]}_{ij}&=
\left(
\begin{matrix}
(A_{n})_{ij}&(B_{n})_{ij}\\[5pt]
(C_{n})_{ij}&(D_{n})_{ij}\\
\end{matrix}
\right),\ \ i,j=1,2,\\[4pt]
H^{[n]}&=
\left(
\begin{matrix}
\lambda_{1}^{0}\varphi_{1,1}&\lambda_{2}^{0}\varphi_{1,2}&\cdots&\lambda_{2n}^{0}\varphi_{1,2n-1}&\lambda_{2n}^{0}\varphi_{1,2n}\\[5pt]
\lambda_{1}^{0}\varphi_{2,1}&\lambda_{2}^{0}\varphi_{2,2}&\cdots&\lambda_{2n}^{0}\varphi_{2,2n-1}&\lambda_{2n}^{0}\varphi_{2,2n}\\[5pt]
\vdots&\vdots&\ddots&\vdots&\vdots\\[5pt]
\lambda_{1}^{n-1}\varphi_{1,1}&\lambda_{2}^{n-1}\varphi_{1,2}&\cdots&\lambda_{2n}^{n-1}\varphi_{1,2n-1}&\lambda_{2n}^{n-1}\varphi_{1,2n}\\[5pt]
\lambda_{1}^{n-1}\varphi_{2,1}&\lambda_{2}^{n-1}\varphi_{2,2}&\cdots&\lambda_{2n}^{n-1}\varphi_{2,2n-1}&\lambda_{2n}^{n-1}\varphi_{2,2n}\\
\end{matrix}
\right),\\[4pt]
(A_{n})_{ij}&=
\begin{cases}
\left(
\begin{matrix}
\lambda^{0}&0&\lambda&0&\cdots&\lambda^{n-1}&0
\end{matrix}
\right),\ \ j=1,\\[5pt]
\left(
\begin{matrix}
0&\lambda^{0}&0&\lambda&\cdots&0&\lambda^{n-1}
\end{matrix}
\right),\ \ j=2,
\end{cases}
(B_{n})_{ij}=
\begin{cases}
\left(
\begin{matrix}
\lambda^{n}
\end{matrix}
\right),\ \ i=j,\\[5pt]
\left(
\begin{matrix}
\ 0\ \
\end{matrix}
\right),\ \ i\neq j,
\end{cases}\\[4pt]
(C_{n})_{ij}&=(H^{[n]})^{\mathrm{T}},
\ \ \
(D_{n})_{ij}=
\left(
\begin{matrix}
\lambda_{1}^{n}\varphi_{i,1}&\lambda_{2}^{n}\varphi_{i,2}&\cdots&\lambda_{2n-1}^{n}\varphi_{i,2n-1}&\lambda_{2n}^{n}\varphi_{i,2n}
\end{matrix}
\right)^{\mathrm{T}}.
\end{align*}
When $\sigma=1$, there are
\begin{equation*}
\varphi_{2,2n}(x,t)=\dot{\zeta}\varphi_{1,2n-1}^{*}(-x,t),\ \ \varphi_{1,2n}(x,t)=\dot{\zeta}\varphi_{2,2n-1}^{*}(-x,t),\ \dot{\zeta}=\pm 1;
\end{equation*}
when $\sigma=-1$, there are
\begin{equation*}
\varphi_{2,2n}(x,t)=-\ddot{\zeta}\varphi_{1,2n-1}^{*}(-x,t),\ \ \varphi_{1,2n}(x,t)=\ddot{\zeta}\varphi_{2,2n-1}^{*}(-x,t),\ \ddot{\zeta}=\pm 1.
\end{equation*}
\end{theorem}
According to the above mentioned Darboux matrix $T_{n}$, we can get the relationships between the new solutions $E^{'},\ p^{'},\ \eta^{'}$ and the old solutions $E,\ p,\ \eta$ of Eq. \eqref{2.17}
\begin{equation}\label{3.15}
E^{[n]}(x,t)=E(x,t)+2i(S^{[n]}_{n-1})_{12},\ \ \ \ \ \ p^{[n]}=\frac{\tilde{p}^{[n]}}{\Delta^{[n]}},\ \ \ \ \ \ \eta^{[n]}=\frac{\tilde{\eta}^{[n]}}{\Delta^{[n]}},\ \ \ \ \ \ \ \
\end{equation}
where $\Delta^{[n]}=(T_{n})_{11}(T_{n})_{22}-(T_{n})_{12}(T_{n})_{21},$
\begin{align*}
\tilde{p}^{[n]}(x,t)&=p(x,t)\cdot\big((T_{n})_{11}\big)^{2}-\sigma p^{*}(-x,t)\cdot\big((T_{n})_{12}\big)^{2}+2\sigma\eta(x,t)\cdot(T_{n})_{11}(T_{n})_{12},\\
\tilde{\eta}^{[n]}(x,t)&=\sigma p(x,t)\cdot(T_{n})_{11}(T_{n})_{21}-p^{*}(-x,t)\cdot(T_{n})_{12}(T_{n})_{22}\\
&\ \ \ +\eta(x,t)\cdot\big((T_{n})_{11}(T_{n})_{22}+(T_{n})_{12}(T_{n})_{21}\big).
\end{align*}

\section{Explicit solutions of $PT$-symmetric NH-MB system}

\ \ \ \ In this section, we will combine the Darboux transformation to find various types of solutions of $PT$-symmetric NH-MB system \eqref{2.17} by giving different seed solutions.\\[4pt]
\textbf{Case $1$: $E=0,\ p=0,\ \eta=x.$}\\

Then $\varphi=(\varphi_{1},\varphi_{2})^{\mathrm{T}}$ satisfies
\begin{equation}\label{4.1}
\varphi_{t}=\left(
\begin{array}{cc}
-i\lambda & 0\\[6pt]
0 & i\lambda\\
\end{array}
\right)\varphi,
\end{equation}
\begin{equation}\label{4.2}
\varphi_{x}=\left(
\begin{array}{cc}
-4\varepsilon_{2}\lambda^{3}+\varepsilon_{1}\lambda^{2}+\frac{ix}{\lambda+\omega}&0\\[6pt]
0&4\varepsilon_{2}\lambda^{3}-\varepsilon_{1}\lambda^{2}-\frac{ix}{\lambda+\omega}\\
\end{array}
\right)\varphi.
\end{equation}
Take a set of special solution of Eqs.~\eqref{4.1} and \eqref{4.2}, in the form shown below
\begin{equation}\label{4.3}
\begin{split}
\varphi=&
\left(
\begin{array}{cc}
\varphi_{1}\\[6pt]
\varphi_{2}\\
\end{array}
\right)
=
\left(
\begin{array}{cc}
e^{-i\lambda t+(-4\varepsilon_{2}\lambda^{3}+\varepsilon_{1}\lambda^{2})x+\frac{i}{2(\lambda+\omega)}x^{2}}\\[6pt]
e^{i\lambda t+(4\varepsilon_{2}\lambda^{3}-\varepsilon_{1}\lambda^{2})x-\frac{i}{2(\lambda+\omega)}x^{2}}\\
\end{array}
\right).
\end{split}
\end{equation}
We assume that $\lambda=\mu_{1}+i\mu_{2}$, there is a relationship between $\mu_{1},\ \mu_{2}$ and $\varepsilon_{1},\ \varepsilon_{2}$
\begin{equation}\label{4.4}
\varepsilon_{1}\mu_{1}\mu_{2}=2\varepsilon_{2}(3\mu_{1}^{2}\mu_{2}^{2}-\mu_{2}^{3}),\ \ \mu_{1}\mu_{2}\neq 0.
\end{equation}
By substituting \eqref{4.3} and the seed solutions into Eqs.~\eqref{3.11}$-$\eqref{3.13} and combining \eqref{4.4}, we can obtain the solutions of Eq.~\eqref{2.17}.\\[4pt]
\textbf{(i) defocusing $PT$-symmetric NH-MB system ($\sigma=1$})\\
\begin{align}\label{4.5}
E^{'}&=-2\mu_{2}\cdot e^{f_{1}}\cdot csch(f_{2}),\\
p^{'}&=\frac{-2i\mu_{2}\cdot e^{f_{1}}}{(\omega+\mu_{1})^{2}+\mu_{2}^{2}}\cdot\left((\omega+\mu_{1})\cdot csch(f_{2})+i\mu_{2}\frac{csch^{2}(f_{2})}{sech(f_{2})}\right)\cdot x,\\
\eta^{'}&=\left(1+\frac{2\mu_{2}^{2}}{(\omega+\mu_{1})^{2}+\mu_{2}^{2}}\cdot csch^{2}(f_{2})\right)\cdot x,
\end{align}
where
\begin{align*}
f_{1}&=-2i\mu_{1}t+\frac{i(\omega+\mu_{1})\cdot x^{2}}{(\omega+\mu_{1})^{2}+\mu_{2}^{2}}-2x\big(4\varepsilon_{2}(\mu_{1}^{3}-3\mu_{1}\mu_{2}^{2})-\varepsilon_{1}(\mu_{1}^{2}-\mu_{2}^{2}) \big),\\
f_{2}&=2\mu_{2}t+\frac{\mu_{2}\cdot x^{2}}{(\omega+\mu_{1})^{2}+\mu_{2}^{2}}.
\end{align*}
Obviously, by analyzing the forms of the solutions $E^{'},\ p^{'},\ \eta^{'},$ we can find that this is a set of singular solutions when $f_{2}=0$, and the solutions $E^{'},\ p^{'},\ \eta^{'}$ develop singularity at $t=\frac{x^{2}}{2\big((\omega+\mu_{1})^{2}+\mu_{2}^{2}\big)}.$                              \\[4pt]
\textbf{(ii) focusing $PT$-symmetric NH-MB system ($\sigma=-1$})\\
\begin{align}\label{4.8}
E^{'}&=2\mu_{2}\cdot e^{f_{1}}\cdot sech(f_{2}),\\
p^{'}&=\frac{-2i\mu_{2}\cdot e^{f_{1}}}{(\omega+\mu_{1})^{2}+\mu_{2}^{2}}\cdot\left((\omega+\mu_{1})\cdot sech(f_{2})+i\mu_{2}\frac{sech^{2}(f_{2})}{csch(f_{2})}\right)\cdot x,\\
\eta^{'}&=\left(1-\frac{2\mu_{2}^{2}}{(\omega+\mu_{1})^{2}+\mu_{2}^{2}}\cdot sech^{2}(f_{2})\right)\cdot x,
\end{align}
where $f_{1}$ and $f_{2}$ are consistent with (i).

We choose $\varepsilon_{2}=-0.1,\ \mu_{1}=-0.5,\ \mu_{2}=0.5,\ \omega=1,$ then the pictorial representations of the solutions of the focusing $PT$-symmetric NH-MB system are shown in Figs.~$1-3$.
\begin{figure}[H]
\centering
\includegraphics[height=5.5cm,width=5.5cm]{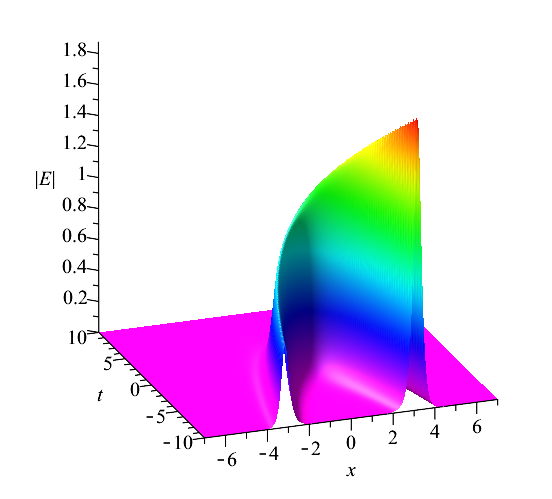}\ \ \ \ \ \ \ \ \ \ \ \
\includegraphics[height=4cm,width=4cm]{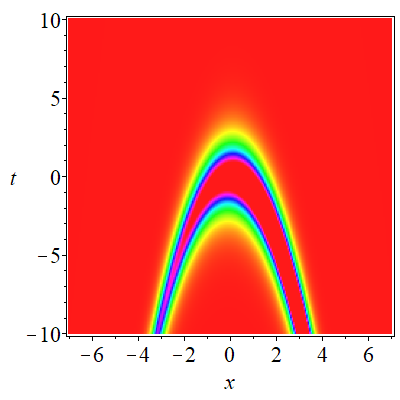}
\caption{\small\footnotesize Solution $E$ of the focusing $PT$-symmetric NH-MB system}
\end{figure}
\begin{figure}[H]
\centering
\includegraphics[height=5.5cm,width=5.5cm]{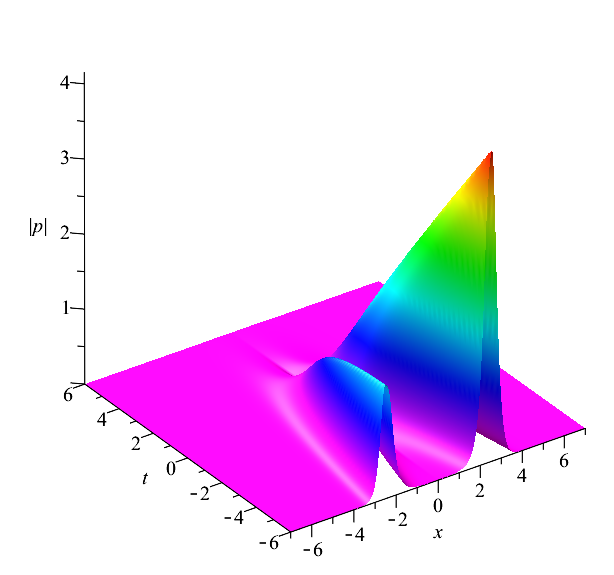}\ \ \ \ \ \ \ \ \ \ \ \
\includegraphics[height=4cm,width=4cm]{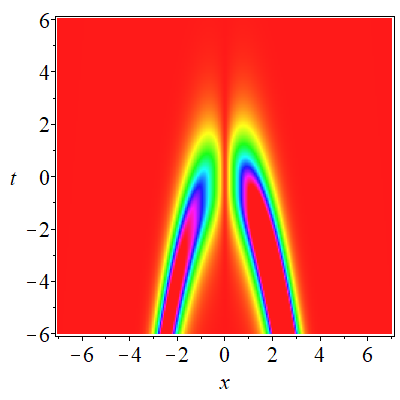}
\caption{\small\footnotesize Solution $p$ of the focusing $PT$-symmetric NH-MB system}
\end{figure}
\begin{figure}[H]
\centering
\includegraphics[height=5.5cm,width=5.5cm]{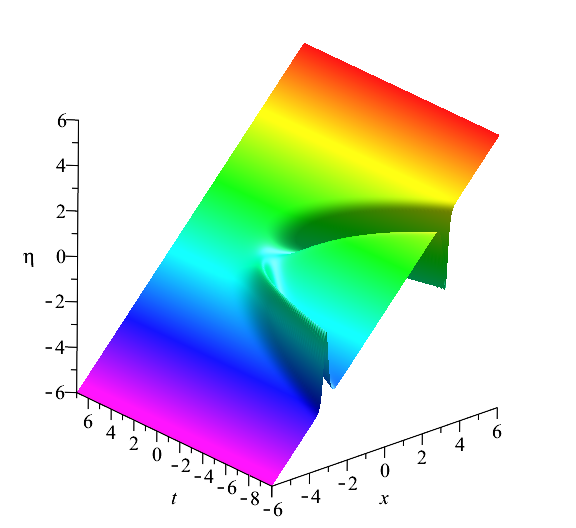}\ \ \ \ \ \ \ \ \ \ \ \
\includegraphics[height=4cm,width=4cm]{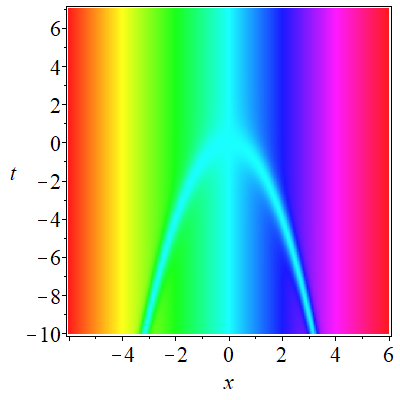}
\caption{\small\footnotesize Solution $\eta$ of the focusing $PT$-symmetric NH-MB system}
\end{figure}
By combining the two forms of diagrams, we can see that all of them are roughly arched. From the density diagram in Fig.~$1$, solution $E$ is a standard arch, but from the first diagram, the right side is obviously higher than the left side; In Fig.~$2$, we can observe that the whole thing of solution $p$ is also an arch, but from its density diagram, the left and right parts will never intersect, which is the difference between solutions $E$ and $p$; In Fig.~$3$, the density diagram of solution $\eta$ along the $x=0$ symmetric, and it can be seen from the first diagram in Fig.~$3$ that the wave is bright on the left and dark on the right.\\
\begin{figure}[H]
\centering
\includegraphics[height=3.3cm,width=3.3cm]{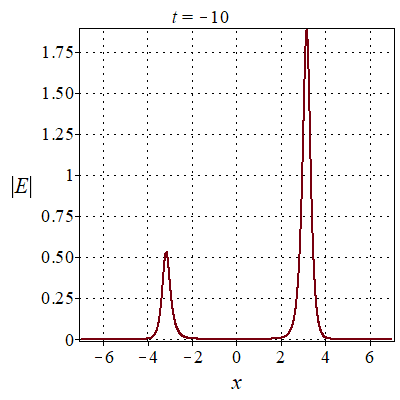}\ \ \
\includegraphics[height=3.3cm,width=3.3cm]{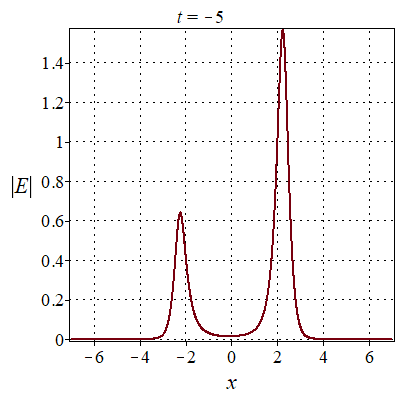}\ \ \
\includegraphics[height=3.3cm,width=3.3cm]{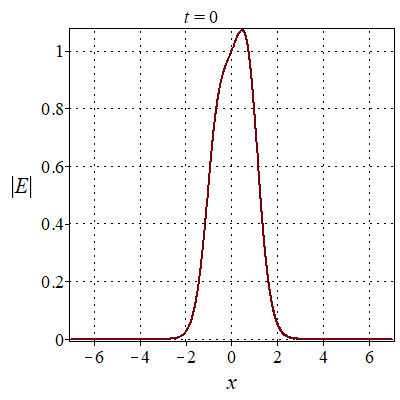}\ \ \
\includegraphics[height=3.3cm,width=3.3cm]{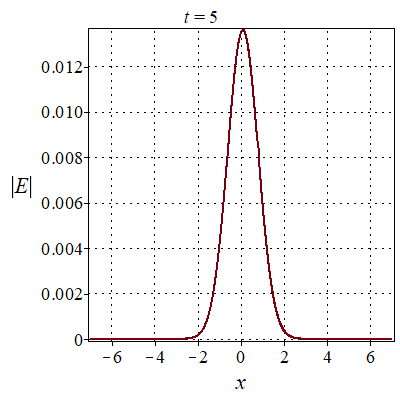}
\caption{Evolution of solution $E$}
\end{figure}
\begin{figure}[H]
\centering
\includegraphics[height=3.3cm,width=3.3cm]{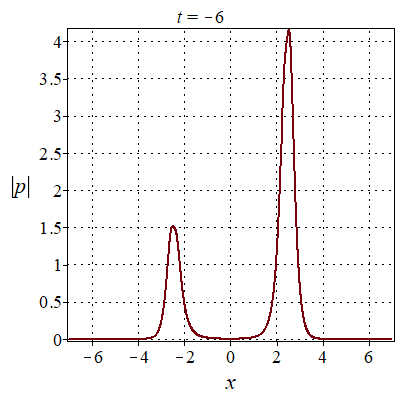}\ \ \
\includegraphics[height=3.3cm,width=3.3cm]{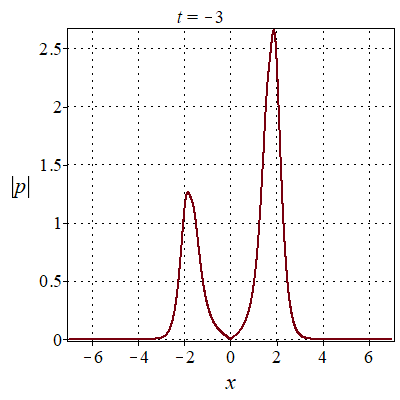}\ \ \
\includegraphics[height=3.3cm,width=3.3cm]{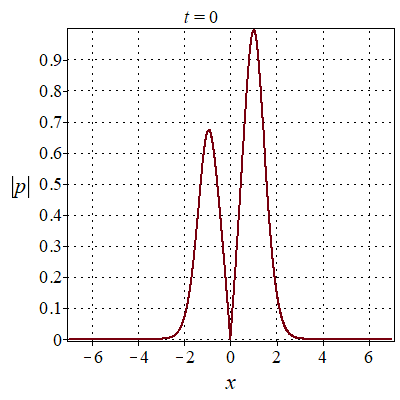}\ \ \
\includegraphics[height=3.3cm,width=3.3cm]{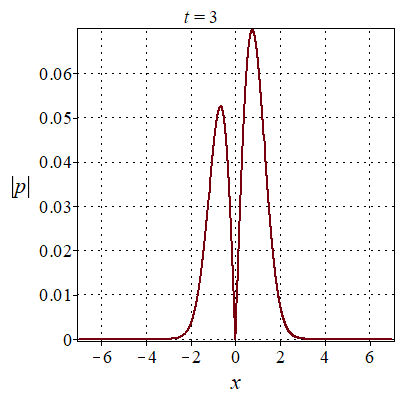}
\caption{Evolution of solution $p$}
\end{figure}
By observing Fig.~$4$ and Fig.~$5$, we can see that two waves of solutions $E$ and $p$ are approaching each other. The difference is that the two waves of solution $E$ eventually merge into one wave, while the two waves of solution $p$ go forward side by side. The height of their peaks  is decreasing over time, and it's not hard to see that they will eventually disappear.\\[4pt]
\textbf{Case $2$: $E=de^{ax+ibt},\ p=ce^{ax+ibt},\ \eta=0.$}

Take the above seed solutions into Eq.~\eqref{2.17}, we can derive the relations
\begin{equation}\label{4.11}
b=2\omega,\ \ a=4\varepsilon_{2}\omega(2\omega^{2}+3\sigma d^{2})+\varepsilon_{1}(2\omega^{2}+\sigma d^{2})+\frac{2\sigma c}{d}.
\end{equation}
Then the linear Eqs.~\eqref{1.1} and \eqref{1.2} becomes
\begin{equation}\label{4.12}
\varphi_{t}=\left(
\begin{array}{cc}
-i\lambda & de^{ax+2i\omega t}\\[6pt]
\sigma de^{-ax-2i\omega t} & i\lambda\\
\end{array}
\right)\varphi,
\ \ \ \varphi_{x}=\left(
\begin{array}{cc}
\hat{N_{11}} & \hat{N_{12}}\\[6pt]
\hat{N_{21}} & \hat{N_{22}}\\
\end{array}
\right)\varphi,
\end{equation}
where
\begin{align*}
\hat{N_{11}}&=-\hat{N_{22}}=-4\varepsilon_{2}\lambda^{3}+\varepsilon_{1}\lambda^{2}-2\sigma\varepsilon_{2}d^{2}\lambda+\sigma(4\varepsilon_{2}\omega +\frac{1}{2}\varepsilon_{1})d^{2},\\
\hat{N_{12}}&=\Big(-4i\varepsilon_{2}\lambda^{2}+(4i\varepsilon_{2}\omega+i\varepsilon_{1})\lambda-2i\varepsilon_{2}(2\omega^{2}+\sigma d^{2})-i\varepsilon_{1}\omega- \frac{i\sigma c}{d(\lambda+\omega)}\Big)\cdot de^{ax+2i\omega t},\\
\hat{N_{21}}&=\sigma\hat{N_{12}}\cdot e^{-2(ax+2i\omega t)}.
\end{align*}
Take a special solution of linear system \eqref{4.12} with the form of
\begin{equation}\label{4.13}
\begin{split}
\varphi=&
\left(
\begin{array}{cc}
\varphi_{1}\\[6pt]
\varphi_{2}\\
\end{array}
\right)
=
\left(
\begin{array}{cc}
e^{ax+2i\omega t}\cdot(e^{-(g_{1}+\gamma_{1}g_{2})x-(i\omega-\frac{R}{2})t}+e^{-(g_{1}+\gamma_{2}g_{2})x-(i\omega+\frac{R}{2})t})\\[6pt]
\frac{1}{d}\cdot(\gamma_{2}e^{-(g_{1}+\gamma_{1}g_{2})x-(i\omega-\frac{R}{2})t}+\gamma_{1}e^{-(g_{1}+\gamma_{2}g_{2})x-(i\omega+\frac{R}{2})t})\\
\end{array}
\right),
\end{split}
\end{equation}
where
\begin{align*}
R&=2\sqrt{\sigma d^{2}-(\lambda+\omega)^{2}},\\
g_{1}&=\hat{N_{11}},\ \ \ g_{2}=\frac{\hat{N_{12}}}{d}\cdot e^{-ax-2i\omega t},\\
\gamma_{1}&=i(\lambda+\omega)-\frac{R}{2},\ \ \ \gamma_{2}=i(\lambda+\omega)+\frac{R}{2}.\\
\end{align*}
We take $\lambda=\mu_{1}+i\mu_{2}$, then $\mu_{1},\ \mu_{1},\ \varepsilon_{1},\ \varepsilon_{2}$ and $\omega$ need meeting the constraints
\begin{equation*}
\varepsilon_{1}=\frac{\varepsilon_{2}(4\mu_{1}^{2}+\sigma d^{2})}{\mu_{1}},\ \ \mu_{1}=\varsigma \mu_{2}=-\omega,\ \varsigma=\pm 1.
\end{equation*}
In particular, for $\sigma=-1$, $\mu_{2}^{2}>d^{2}$.

We do not represent the expressions of $p$ and $\eta$ here as their expressions are too long.\\[4pt]
\textbf{(i) defocusing $PT$-symmetric NH-MB system ($\sigma=1$})

Let's take $E$ as an example, which can be represented as
\begin{equation*}
E=de^{ax+2i\omega t}\cdot\left(1-4\varsigma \mu_{1}\cdot\frac{\gamma_{2}e^{Rt}+\gamma_{1}e^{-Rt}+\gamma_{1}e^{g_{2}Rx}+\gamma_{2}e^{-g_{2}Rx}}
{(d^{2}-\gamma_{2}^{2})e^{Rt}+(d^{2}-\gamma_{1}^{2})e^{-Rt}+2d^{2}(e^{g_{2}Rx}+e^{-g_{2}Rx})}\right).
\end{equation*}
Solutions of the defocusing $PT$-symmetric NH-MB system are obtained as Figs.~$6-8$.
\begin{figure}[H]
\centering
\includegraphics[height=5.5cm,width=5.5cm]{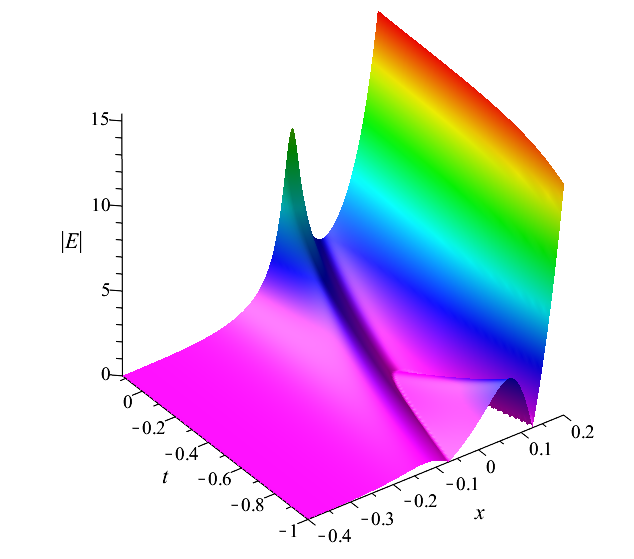}\ \ \ \ \ \ \ \ \ \ \ \
\includegraphics[height=4cm,width=4cm]{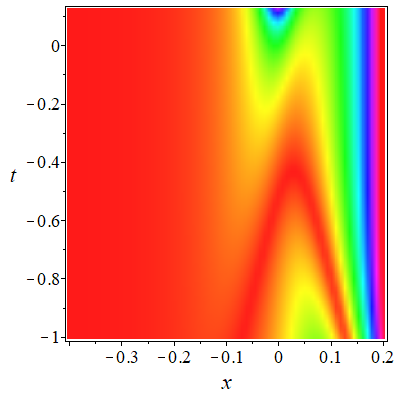}
\caption{\small\footnotesize Solution $E$ of the defocusing $PT$-symmetric NH-MB system}
\end{figure}
\begin{figure}[H]
\centering
\includegraphics[height=5.5cm,width=5.5cm]{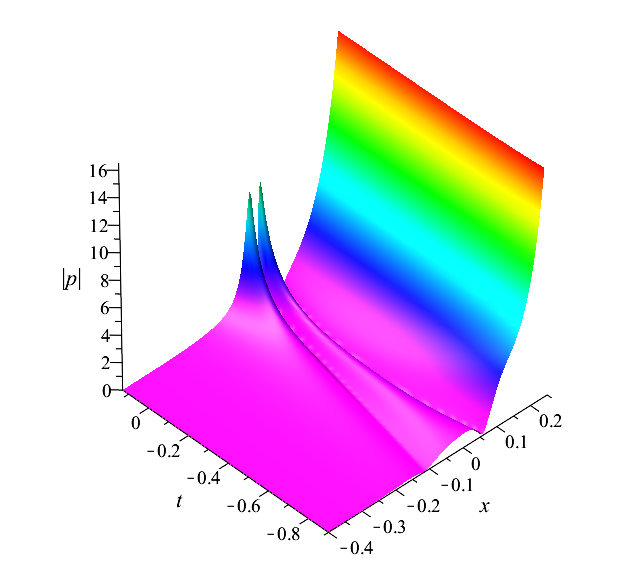}\ \ \ \ \ \ \ \ \ \ \ \
\includegraphics[height=4cm,width=4cm]{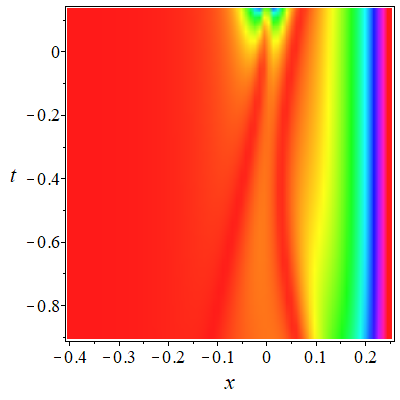}
\caption{\small\footnotesize Solution $p$ of the defocusing $PT$-symmetric NH-MB system}
\end{figure}
\begin{figure}[H]
\centering
\includegraphics[height=5.5cm,width=5.5cm]{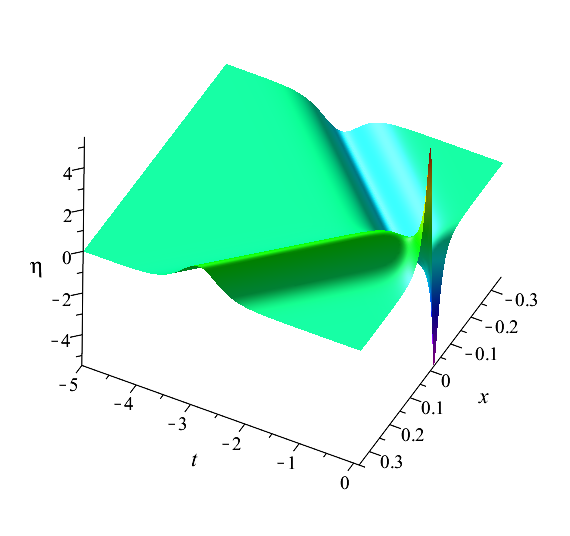}\ \ \ \ \ \ \ \ \ \ \ \
\includegraphics[height=4cm,width=4cm]{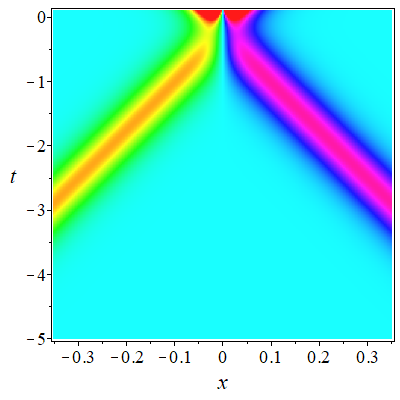}
\caption{\small\footnotesize Solution $\eta$ of the defocusing $PT$-symmetric NH-MB system}
\end{figure}
The corresponding parameters in above diagrams are $d=2,\ c=-1,\ \varepsilon_{2}=2,\ \mu_{1}=-1,\ \varsigma=1.$ From the Fig.~$6$ and Fig.~$7$, we can observe that the changing forms of waves look very similar, in this process from negative infinity to positive infinity along the $x$-axis, the waves of them start from zero, and finally round towards positive infinity. For the wave of solution $\eta$ in Fig.~$8$, when $x$ rounds towards infinity, it always tends to zero. Fig.~$9$ and Fig.~$10$ are the time evolution diagrams of solutions $E$ and $p$, respectively.\\
\begin{figure}[H]
\centering
\includegraphics[height=3.3cm,width=3.3cm]{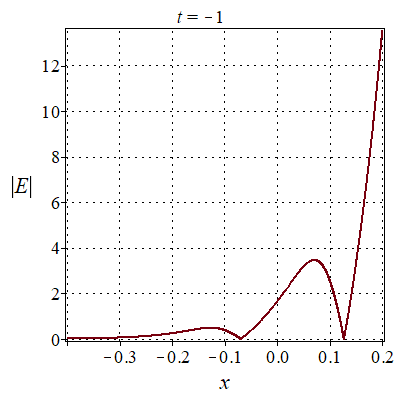}\ \ \
\includegraphics[height=3.3cm,width=3.3cm]{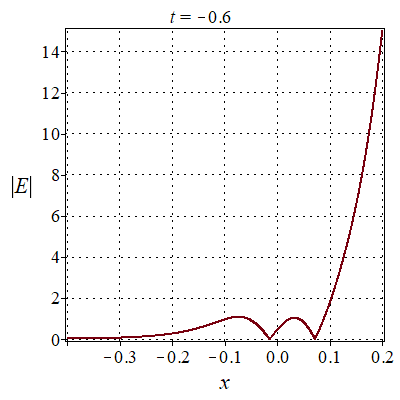}\ \ \
\includegraphics[height=3.3cm,width=3.3cm]{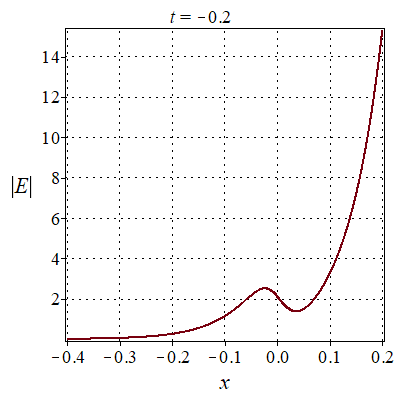}\ \ \
\includegraphics[height=3.3cm,width=3.3cm]{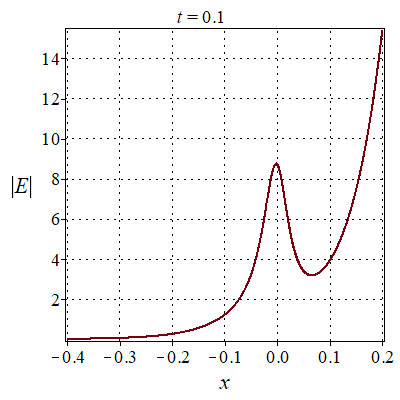}
\caption{Evolution of solution $E$}
\end{figure}
\begin{figure}[H]
\centering
\includegraphics[height=3.3cm,width=3.3cm]{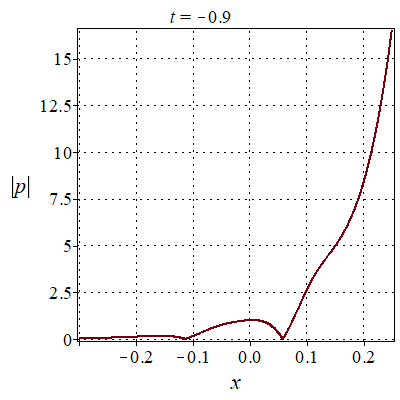}\ \ \
\includegraphics[height=3.3cm,width=3.3cm]{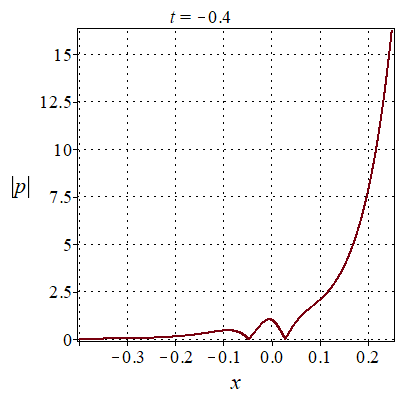}\ \ \
\includegraphics[height=3.3cm,width=3.3cm]{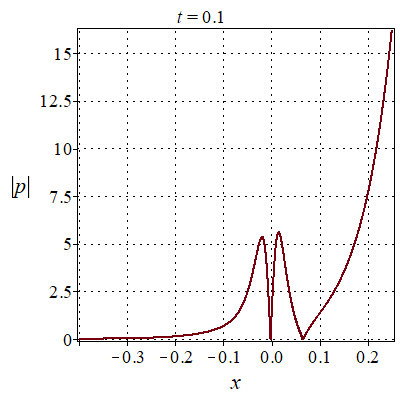}\ \ \
\includegraphics[height=3.3cm,width=3.3cm]{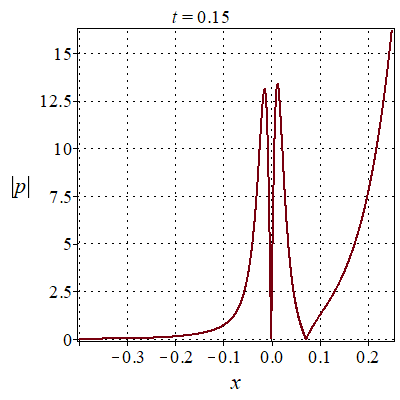}
\caption{Evolution of solution $p$}
\end{figure}
From Fig.~$9$ and Fig.~$10$, we can see that the two waves of solution $E$ are still combined into one wave eventually, while the two waves of solution $p$ are close to each other but do not blend, which is the same as their variation forms in the solutions of the focusing $PT$-symmetric NH-MB system. The difference is that the heights of their wave peaks no longer decrease with time, but increase. In addition, the peak of the first wave in Fig.~$9$ is gradually higher than that of the second wave over time.\\[4pt]
\textbf{(ii) focusing $PT$-symmetric NH-MB system ($\sigma=-1$})

Similar to (i), we can obtain
\begin{equation*}
E=de^{ax+2i\omega t}\cdot\left(1+4\varsigma \mu_{1}\cdot\frac{\gamma_{2}e^{Rt}+\gamma_{1}e^{-Rt}+\gamma_{1}e^{g_{2}Rx}+\gamma_{2}e^{-g_{2}Rx}}
{(d^{2}+\gamma_{2}^{2})e^{Rt}+(d^{2}+\gamma_{1}^{2})e^{-Rt}+2d^{2}(e^{g_{2}Rx}+e^{-g_{2}Rx})}\right).
\end{equation*}
\begin{figure}[H]
\centering
\includegraphics[height=5.5cm,width=5.5cm]{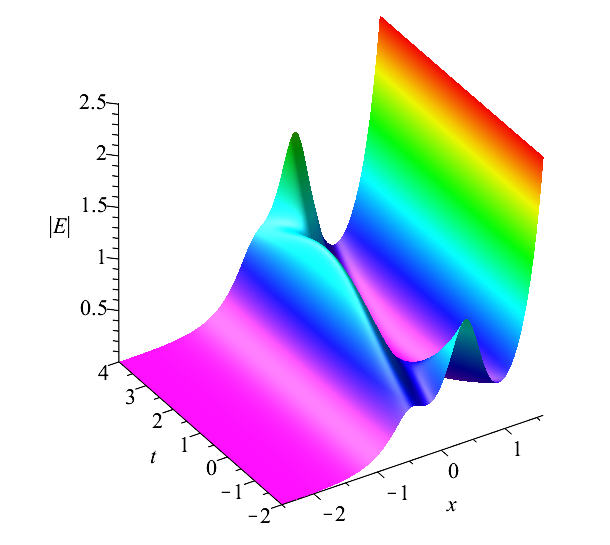}\ \ \ \ \ \ \ \ \ \ \ \
\includegraphics[height=4cm,width=4cm]{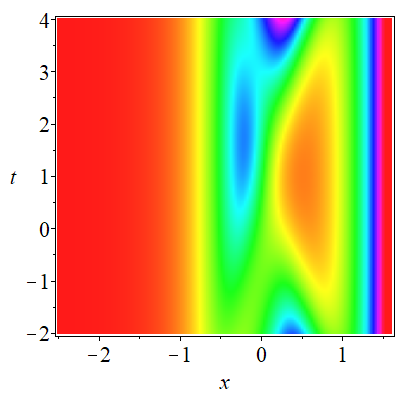}
\caption{\small\footnotesize Solution $E$ of the focusing $PT$-symmetric NH-MB system}
\end{figure}
\begin{figure}[H]
\centering
\includegraphics[height=5.5cm,width=5.5cm]{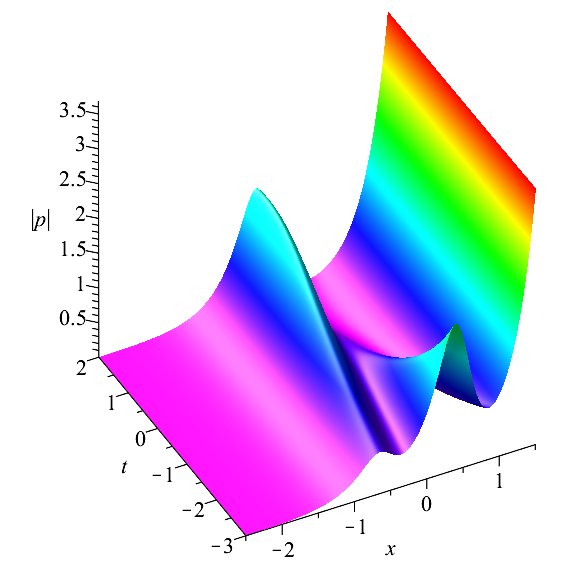}\ \ \ \ \ \ \ \ \ \ \ \
\includegraphics[height=4cm,width=4cm]{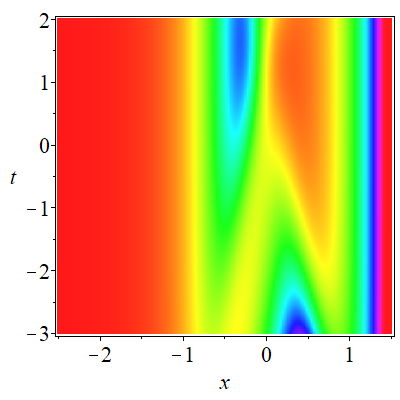}
\caption{\small\footnotesize Solution $p$ of the focusing $PT$-symmetric NH-MB system}
\end{figure}
\begin{figure}[H]
\centering
\includegraphics[height=5.5cm,width=5.5cm]{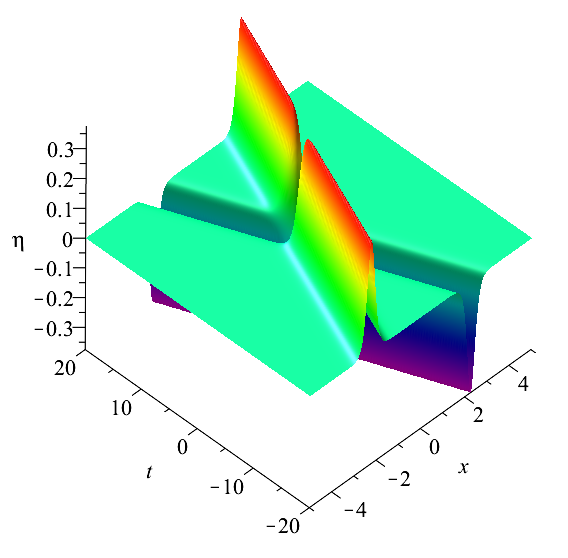}\ \ \ \ \ \ \ \ \ \ \ \
\includegraphics[height=4cm,width=4cm]{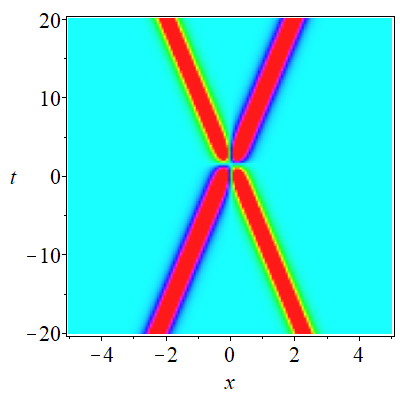}
\caption{\small\footnotesize Solution $\eta$ of the focusing $PT$-symmetric NH-MB system}
\end{figure}
Solutions are shown in Figs.~$11-13$ with different values of the parameters $d,\ c,\ \varepsilon_{2},\ \mu_{1}$ and $\varsigma$, respectively. Here we take $d=0.25,\ c=-0.5,\ \varepsilon_{2}=-3,\ \mu_{1}=0.5,\ \varsigma=1.$ From Figs.~$11$-$13$, we can observe that the changing forms of waves along the $x$-axis are consistent with that of (i). It is worth noting that the propagation form of solution $\eta$ along the $t$-axis is interesting. From Fig.~$13$, we can see that the wave on the left is bright and the wave on the right is dark. But the propagation paths of the two waves are reversed at some point, that is, the wave on the left becomes dark and the wave on the right becomes bright, and then keep their shapes and continue to propagate.

\section{Discussions of the reverse space-time NH-MB system}
\ \ \ \ In this section, we also study the reverse space-time NH-MB system \eqref{2.20} by using the Darboux transformation, and discuss the solutions of the Eq.~\eqref{2.20}. Similarly, we analyze Eqs.~\eqref{3.5} and \eqref{3.6} respectively. Here, we substitute Eq.~\eqref{3.10} into Eqs.~\eqref{3.5} and \eqref{3.6}, combine the symmetric form \eqref{2.19} and Eq.~\eqref{3.1} for calculation, and then compare the coefficients of each power of $\lambda$, we could find that when $\sigma=1$, there are
\begin{equation*}
s_{11}(x,t)=\dot{\xi}s_{22}(-x,-t),\ s_{12}(x,t)=-\dot{\xi}s_{21}(-x,-t),\ a_{11}(x,t)=\dot{\xi}a_{22}(-x,-t),\ \ \dot{\xi}=\pm 1;
\end{equation*}
when $\sigma=-1$, there are
\begin{equation*}
s_{11}(x,t)=\ddot{\xi}s_{22}(-x,-t),\ s_{12}(x,t)=\ddot{\xi}s_{21}(-x,-t),\ a_{11}(x,t)=\ddot{\xi}a_{22}(-x,-t),\ \ \ddot{\xi}=\pm 1.
\end{equation*}
Obviously, they are different from the relations obtained in the $PT$-symmetric NH-MB system \eqref{2.17}, and the direct reason was the different selection of the symmetric form. Then we obtain
\begin{equation}\label{5.1}
a_{22}E^{'}=a_{11}E-2is_{12},
\end{equation}
\begin{equation*}
a_{12}=a_{21}=0,\ \ a_{11t}=a_{22t}=0,
\end{equation*}
\begin{equation}\label{5.2}
N^{'}_{-1}=(S+\omega A)N_{-1}(S+\omega A)^{-1},
\end{equation}
where
\begin{equation*}
N_{-1}=\left(
\begin{array}{cc}
\eta(x,t)&-\sigma p(x,t)\\[6pt]
-p(-x,-t)&-\eta(x,t)\\
\end{array}
\right).
\end{equation*}
Similarly, we assume that the matrix $S$ is in the form of formula \eqref{3.10}. By calculation, we find that if we take $A=I$, then $S=\lambda_{1}\sigma_{3}$. At this time, $S$ is meaningless, so we take $A=\sigma_{3}$ here, and then we have
\begin{align*}
\sigma&=1: s_{11}(x,t)=-s_{22}(-x,-t),\ s_{12}(x,t)=s_{21}(-x,-t);\\
\sigma&=-1: s_{11}(x,t)=-s_{22}(-x,-t),\ s_{12}(x,t)=-s_{21}(-x,-t).
\end{align*}
Combined with the relationships between $s_{11},\ s_{12}$ and $s_{22},\ s_{21}$, through a series of calculations, there are
\begin{align*}
\lambda_{1}&=-\lambda_{2},\\
\varphi_{11}(x,t)&=\zeta\varphi_{22}(-x,-t),\\
\varphi_{12}(x,t)&=-\sigma\zeta\varphi_{21}(-x,-t),\ \zeta=\pm 1,
\end{align*}
therefore
\begin{equation}\label{5.3}
S=\frac{\lambda_{1}}{\Delta}
\cdot\left(
\begin{array}{cc}
\varphi_{1,1}\varphi_{1,1}(-x,-t)-\sigma\varphi_{2,1}\varphi_{2,1}(-x,-t)&2\sigma\varphi_{1,1}\varphi_{2,1}(-x,-t)\\[6pt]
2\varphi_{1,1}(-x,-t)\varphi_{2,1}&-\varphi_{1,1}\varphi_{1,1}(-x,-t)+\sigma\varphi_{2,1}\varphi_{2,1}(-x,-t)\\
\end{array}
\right),\\[4pt]
\end{equation}
where $\Delta=\varphi_{1,1}(x,t)\varphi_{1,1}(-x,-t)+\sigma\varphi_{2,1}(x,t)\varphi_{2,1}(-x,-t)$.

Then we substitute \eqref{5.3} into Eqs.~\eqref{5.1} and \eqref{5.2} to get the relationships between new solutions $E^{'},\ p^{'},\ \eta^{'}$ and old solutions $E,\ p,\ \eta$ of Eq.~\eqref{2.20}, as shown below
\begin{equation}\label{5.4}
E^{'}(x,t)=-E(x,t)+\frac{4i\sigma\lambda_{1}\varphi_{1,1}(x,t)\varphi_{2,1}(-x,-t)}{\varphi_{1,1}(x,t)\varphi_{1,1}(-x,-t)+\sigma\varphi_{2,1}(x,t)\varphi_{2,1}(-x,-t)},\ \ \ \ \ \ \ \ \ \ \ \ \ \
\end{equation}
\begin{equation}\label{5.5}
\begin{split}
\ \ \ p^{'}(x,t)=&\frac{1}{\Delta^{'}}
\Big[-\sigma p(-x,-t)\cdot\big(2\lambda_{1}\varphi_{1,1}(x,t)\varphi_{2,1}(-x,-t)\big)^{2}\\
&+p(x,t)\cdot\big((\omega+\lambda_{1})\varphi_{1,1}(x,t)\varphi_{1,1}(-x,-t)+\sigma(\omega-\lambda_{1})\varphi_{2,1}(x,t)
\varphi_{2,1}(-x,-t)\big)^{2}\\
&+4\eta(x,t)\cdot
\lambda_{1}\varphi_{1,1}(x,t)\varphi_{2,1}(-x,-t)\cdot\big((\omega+\lambda_{1})\varphi_{1,1}(x,t)\varphi_{1,1}(-x,-t)\\
&+\sigma(\omega-\lambda_{1})\varphi_{2,1}(x,t)\varphi_{2,1}(-x,-t)\big)\Big],\\[-15pt]
\end{split}
\end{equation}
\begin{equation}\label{5.6}
\begin{split}
\eta^{'}(x,t)=&\frac{1}{\Delta^{'}}
\Big[2\sigma p(-x,-t)\cdot\lambda_{1}\varphi_{1,1}(x,t)\varphi_{2,1}(-x,-t)\\
&\cdot\big((\omega+\lambda_{1})\varphi_{1,1}(x,t)\varphi_{1,1}(-x,-t)+\sigma(\omega-\lambda_{1})\varphi_{2,1}(x,t)
\varphi_{2,1}(-x,-t)\big)\\
&+2\sigma p(x,t)\cdot\lambda_{1}\varphi_{1,1}(-x,-t)\varphi_{2,1}(x,t)\\
&\cdot\big((\omega+\lambda_{1})\varphi_{1,1}(x,t)\varphi_{1,1}(-x,-t)+\sigma(\omega-\lambda_{1})\varphi_{2,1}(x,t)
\varphi_{2,1}(-x,-t)\big)\Big]\\
&+\eta(x,t)\cdot\left(1+\sigma\frac{8}{\Delta^{'}}\lambda_{1}^{2}\varphi_{1,1}(x,t)\varphi_{1,1}(-x,-t)\varphi_{2,1}(x,t)\varphi_{2,1}(-x,-t)            \right),\ \ \ \ \ \ \ \ \
\end{split}
\end{equation}
where
\begin{equation*}
\begin{split}
\Delta^{'}=&-\big((\omega+\lambda_{1})\varphi_{1,1}(x,t)\varphi_{1,1}(-x,-t)+\sigma(\omega-\lambda_{1})\varphi_{2,1}(x,t)
\varphi_{2,1}(-x,-t)\big)^{2}\\
&-4\sigma\lambda_{1}^{2}\varphi_{1,1}(x,t)\varphi_{1,1}(-x,-t)\varphi_{2,1}(x,t)\varphi_{2,1}(-x,-t).
\end{split}
\end{equation*}

Similarly, given the seed solutions, we will get the corresponding new solutions of the reverse space-time NH-MB system \eqref{2.20}. Here we only discuss two seed solutions of the same types as in the section $4$. For example, we can take $E=0,\ p=0,\ \eta=1$ or $E=\tilde{d}e^{i(\tilde{a}x+\tilde{b}t)},\ p=i\tilde{c}e^{i(\tilde{a}x+\tilde{b}t)},\ \eta=1$. However, we find that the $\varphi=[\varphi_{1},\varphi_{2}]^{\mathrm{T}}$ of these two kinds of seed solutions are all in the form of $e$-index. Combined with Eqs.~\eqref{5.4}$-$\eqref{5.6}, it is obvious that the new solutions forms finally generated are consistent: both $E$ and $p$ are in the form of $e$-index, while $\eta$ is a constant variable.\\

\section{Conclusions}
\ \ \ \ In order to achieve the compatibility of the five equalities between Eq.~\eqref{2.4} arising from the zero curvature condition for the H-MB system, we exploited various possibility of different combinations between $q(x,t)$ and $r(x,t)$, $p(x,t)$ and $m(x,t)$, including parity, time-space inversion, and complex conjugation. In general, each possibility can correspond to a new type of integrable system. However, due to the limitation of some parameters in the H-MB system \eqref{1.8}, we only get three kinds of integrable systems: the standard H-MB system \eqref{1.8} and two kinds of nonlocal H-MB system. The two nonlocal H-MB systems are $PT$-symmetric NH-MB system \eqref{2.17} and reverse space-time NH-MB system \eqref{2.20}, respectively.

In this paper, the explicit solutions of the focusing and defocusing $PT$-symmetric NH-MB systems \eqref{2.17} are constructed by using the Darboux transformation. In addition, the solutions of the reverse space-time NH-MB system \eqref{2.20} is also discussed briefly. The results show that the defocusing $PT$-symmetric NH-MB system shows singularity when the seed solution is $E=0,\ p=0,\ \eta=x$, and the singularity appears at $t=x^{2}/2\big((\omega+\mu_{1})^{2}+\mu_{2}^{2}\big),$ the singularities of nonlocal systems are also the topic worthy to be discussed\cite{20,21}. For the reverse space-time NH-MB system \eqref{2.20}, we take two different seed solutions respectively, and get the same forms of the new solutions: both $E(x,t)$ and $p(x,t)$ are in the form of $e$-index, and $\eta(x,t)$ is always a constant variable.

For the two NH-MB systems studied in this paper, there are a variety of interesting issues to be explored. Obviously, more specific scenarios can be explored for the above cases and further solutions can be built, for instance by taking different seed function in the Darboux transformation. We also ignore other methods to solve such systems, such as Hirota's direct method, inverse scattering transformation and so on. And we can even investigate whether these nonlocal solutions can be realized experimentally. Furthermore, we can construct the Riemann-Hilbert problems related to the resulting nonlocal systems, as Eq.~\eqref{2.17} and Eq.~\eqref{2.20}.\\

{\bf {Acknowledgements:}}
Chuanzhong Li  is  supported by the National Natural Science Foundation of China under Grant No. 12071237 and K. C. Wong Magna Fund in
Ningbo University.\\

\end{document}